\documentstyle[twoside,fleqn,espcrc2,fps]{article}

\newcommand{\AmS}{{\protect\the\textfont2
  A\kern-.1667em\lower.5ex\hbox{M}\kern-.125emS}}

\title{Method for Extracting the Glueball Wave Function
\thanks{Supported by National Natural Science
  Foundation of China}
}

\author{Jinming Liu, Xiang-Qian Luo, Xiyan Fang,
Shuohong Guo\address{CCAST
     (World Laboratory),  P.O. Box 8730, Beijing 100080, China,\\
       Department of Physics,
       Zhongshan University, Guangzhou 510275, 
       China (Mailing address),\\
       and Center for Computational Physics,
       Zhongshan University, Guangzhou 510275, China}
	Helmut Kr\"oger\address{ D\'epartement 
	de Physique, Universit\'e Laval,
Qu\'ebec, Qu\'ebec G1K 7P4, Canada}
	 Dieter Sch\"utte\address{Institut 
	 f\"ur Theoretische Kernphysik,
Universit\"at Bonn, D-53115 Bonn, Germany}
	Lee Lin\address{Department of Physics, 
	National Chung Hisng University,
Taichung, Taiwan, China}
	      }

\begin{document}

\begin{abstract}
We describe a nonperturbative method for calculating
the QCD vacuum and glueball wave functions, based on an eigenvalue 
equation approach to Hamiltonian lattice gauge theory.
Therefore, one can obtain more physical information than the conventional
simulation methods.
For simplicity, we take the 2+1 dimensional
 U(1) model as an example. The generalization of this method
 to 3+1 dimensional QCD is straightforward.
\end{abstract}

\maketitle

\section{Wave Functions}

To study more preciesely the glueball properties, one should compute
not only the spectroscopy, but also the wave functions. 
Once the wave functions 
are available, the matrix elements relevant 
for the glueball production and decays
(branch ratios) become calculable. 
In some sence, the wave functions \cite{LC,KS} can give more
physical information than the masses themselves.
In this paper, we present a method for such a purpose.

We begin with the vacuum wave function in the form
\begin{eqnarray*}
\mid \Omega>=e^{R(U)}\mid 0>,  
\end{eqnarray*}
where $R(U)$ consists of Wilson loops (clusters) \cite{Green}, and the
state $\mid 0>$ is defined by  $ E_l \mid 0>=0 $.
For the glueballs, we consider here only the
anti-symmetric (under a plane
parity transformation) lowest lying  excited state,
and take the wave function as
\begin{eqnarray*}
\mid \Psi_A >=F^A(U)e^{R(U)}\mid 0>,  
\end{eqnarray*}
where $F^A(U)$ contains various Wilson loops 
with the appropriate symmetry.
It has been shown \cite{GCL} that we can establish 
a truncation scheme, which
preserves the continuum limit, where
the operators R and $F^A$
are expanded
in order of graphs (clusters):
\begin{eqnarray*}
R=R_1 + R_2 + \cdots,  ~~~
F^A=F^A_1+F^A_2+\cdots ~.
\end{eqnarray*}
Here $R_1$ (or $F^A_1$) is the lowest order term in $R$ (or $F^A$), 
and is chosen to be
\begin{eqnarray*}
R_1= r_1 G_1  ,~~~
G_1 \equiv{1\over 2}\sum_x \left[U_p(x)+ h.c. \right],   
\end{eqnarray*}
\begin{eqnarray*}
F^A_1 =a_1 G^A_1, ~~~
G^A_1\equiv{1\over 2} \sum_x\left[U_p(x)-h.c.\right]. 
\end{eqnarray*}
with a coefficient $ r_1 $ (and $a_1$ ) to be determined. 
Higher order clusters
can be produced by solving the eigenvalue equations \cite{GCL,Sch} for
$\mid \Omega>$ and $\mid\Psi_A >$ order by order, from which
we obtain
\begin{eqnarray}
R= r_j G_j , ~~~  F^A = f_j G^A_j.            
\label{1.7}
\end{eqnarray}
Here the repeated index j implies a summation over all the clusters up
to some n-th order. In previous papers \cite{GCL,CLG,FLG,Luo}, 
we have shown that
this method is very efficient 
in obtaining scaling behavior of physical
quantities. First results \cite{Luo} for the glueball masses in QCD 
have been obtained from this method.

\section{Correlation lengths}

In order to obtain the correlation of the states, we first investigate
the continuum limit of the clusters. Expanding $ U_p$ (plaquette), 
$ G_j$, and $G^A_j $  in order of the lattice spacing $a$, we have 
\begin{eqnarray*}
   U_p = e^{-ie\Phi} ,~~~ \Phi = a^2 {\cal F}
+ {a^4\over 24}({\cal D}^2_1 + {\cal D}^2_2 ) {\cal F} + \cdots 
\end{eqnarray*}
Then 
\begin{eqnarray*}
  G_j = 1 - A_j a^4 e^2 {\cal F}^2 
- B_j a^6 e^2 {\cal F} ({\cal D}^2_1 + {\cal D}^2_2){\cal F} +\cdots,
\end{eqnarray*}
\begin{eqnarray*}
  G_j^A = - X_j a^6 e^3 {\cal F}^3 
\end{eqnarray*}
\begin{eqnarray}
- Y_j a^8 e^3 {\cal F}^2 ({\cal D}^2_1 + {\cal D}^2_2){\cal F} +\cdots  
  \label{2.2}
\end{eqnarray}
where $ {\cal F}={\cal F}_{12} $ is the field strength tensor, 
$ {\cal D}_1$ and ${\cal D}_2$ are
the covariant derivatives, and $A_j$ 
and $B_j $ or $ X_j$ and $Y_j $ are constants,
corresponding to the cluster $G_j$ or $G^A_j $ respectively.
The long wave length vacuum wave function is \cite{Arisue,GCL,CLG}
\begin{eqnarray*}
\vert \Omega > \sim e^{-\int d^2 x [\mu_0 {\cal F}^2
+\mu_2 {\cal F} ({\cal D}^2_1 
+ {\cal D}^2_2){\cal F}]}
\end{eqnarray*}
According to Eqs. (\ref{1.7}) and (\ref{2.2}),
\begin{eqnarray*}
  \mu_0 = r_j A_j a^2 e^2, ~~~ \mu_2 = r_j B_j a^4 e^2,
\end{eqnarray*}
from which we obtain the correlation length between
the field strengths of the vacuum
$\xi_v $
\begin{eqnarray*}
   \xi_v = a \sqrt { r_j B_j / r_j A_j }.       
\end{eqnarray*}
Similarly, the long wavelength anti-symmetric 
glueball operator is
\begin{eqnarray*}
   F^A (U) \sim - \int d^2 x [\mu^A_0 {\cal F}^3
	      +\mu^A_2 {\cal F}^2 ({\cal D}^2_1 
	      + {\cal D}^2_2){\cal F}] 
\end{eqnarray*}
and
\begin{eqnarray*}
  \mu^A_0 = f_j X_j a^4 e^3, ~~~ \mu^A_2 = f_j Y_j a^6 e^3.  
\end{eqnarray*}
The correlation length 
in the anti-symmetric glueball state $\xi_A $ is
\begin{eqnarray*}
   \xi_A = a \sqrt { f_j Y_j / f_j X_j }        
\end{eqnarray*}
At this point, it should be noted that the results above are
very general in  the sense 
that the spacial dimension and the gauge
group of the theory are not specified.

For illustration and simplicity, 
we consider here a 2+1 dimensional U(1) model.
It is well known \cite{Poly} that the theory  is confining 
for all non-vanishing coupling constant. 
When $a$ goes to zero, the
glueball mass $Ma$ is expected to decrease exponentially as
\begin{eqnarray*}
M^2 a^2 \sim {c_1 \over g^2} exp\left(-{c_2 \over g^2}\right), 
\end{eqnarray*}
where $c_1$ and $c_2$ are some constants. Hence, in the scaling region
\begin{eqnarray*}
   \mu_0\prime = 2 ~ ln(gr_j A_j)
	 = 2 ~ ln(\mu_0 M c_1^{-\small{1\over 2}}) + c_2 \beta ,
\end{eqnarray*}
\begin{eqnarray*}
  \mu_2\prime = {\small {2\over 3}} ln(g^{-1} r_j B_j)
  = {\small {2\over 3}} ln(\mu_2 M c_1^{-\small {3\over 2}})
		+ c_2 \beta,
\end{eqnarray*}
\begin{eqnarray*}
  \xi_v\prime = ln(\beta r_j B_j  /  r_j A_j )
	      =ln(\xi_v^2 M^2 c_1^{-1}) + c_2 \beta,       
\end{eqnarray*}
\begin{eqnarray*}
\mu^A_0 \prime = {\small {4\over 5}} ln(g^{\small {1\over 2}}f_j X_j)
   ={\small {4\over 5}}
       ln(\mu^A_0 M^{\small{5\over 2}} c_1^{-\small{5\over 4}})
			     + c_2 \beta,          
\end{eqnarray*}
\begin{eqnarray*}
\mu^A_2 \prime = {\small {4\over 9}} ln(g^{\small{-3\over 2}} f_j Y_j)
  = {\small {4\over 9}}
	 ln(\mu^A_2 M^{\small{9\over 2}} c_1^{-\small {9\over 4}})
		+ c_2 \beta,                           
\end{eqnarray*}
\begin{eqnarray*}
  \xi_A \prime = ln(\beta f_j Y_j  / f_j X_j )
	      =ln(\xi_A^2 M^2 c_1^{-1}) + c_2 \beta.       
\end{eqnarray*}
In the continuum limit, $\mu_0$, $\mu_2$, $\mu^A_0$,
$\mu^A_2$, $\xi_v$, $\xi_A$ and $M$  should be
constants, which means the curves of $\mu^\prime_0$, $\mu^\prime_2$,
......
against $\beta$ will be straight line with the expected slope
in the scaling region.

In Figs. 1 and 2, we present the
results for $\mu^\prime_0$ and $\mu^\prime_2$ 
	against $\beta $ from the 2nd to 5th order.
In Fig. 3, we also show the
result for $\mu_0^{A}\prime$
from the 2nd order through 4th order  against  $\beta$.
All the curves
show a nice exponential behavior
and  a clear trend towards convergence.
In particular, the slope of $\mu_0^{\prime}$ is well consistent 
with the spectrum.
For $\mu^\prime_2$,$\mu_0^{A}\prime$, $\mu^\prime_{A2}$, 
$\xi^\prime_v$ and $\xi^\prime_A$, we expect that when the order increases,
their slopes would approach their correct theoretical values.

In conclusion, the method described above has proved to be very useful
for calculating the vacuum and glueball
wave functions and correlation lengths. The calculations in
3+1 dimensional QCD is in progress and the results will be
reported elsewhere.

\begin{figure}[htb]
\fpsxsize=7.5cm
\hspace{-15mm}
\def\fpsangle{270}
\fpsbox[70 90 579 760]{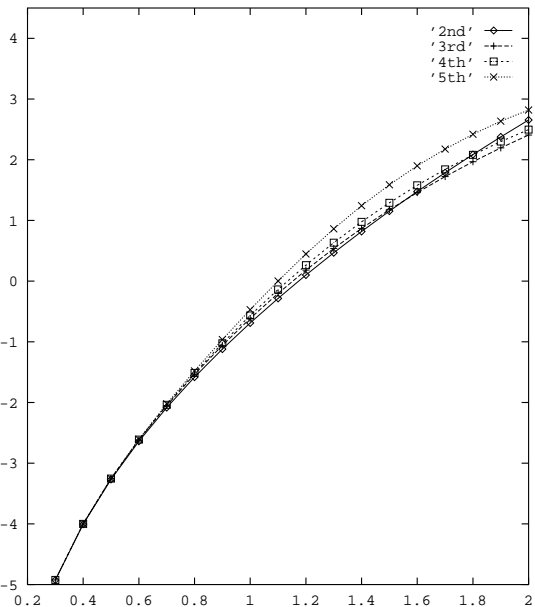}
\vspace{-2mm}
\caption{$\mu_0\prime$ as a function of $\beta$}
\label{fig1}
\end{figure}

\begin{figure}[htb]
\fpsxsize=7.5cm
\hspace{-15mm}
\vspace{-5.1mm}
\def\fpsangle{270}
\fpsbox[70 90 579 760]{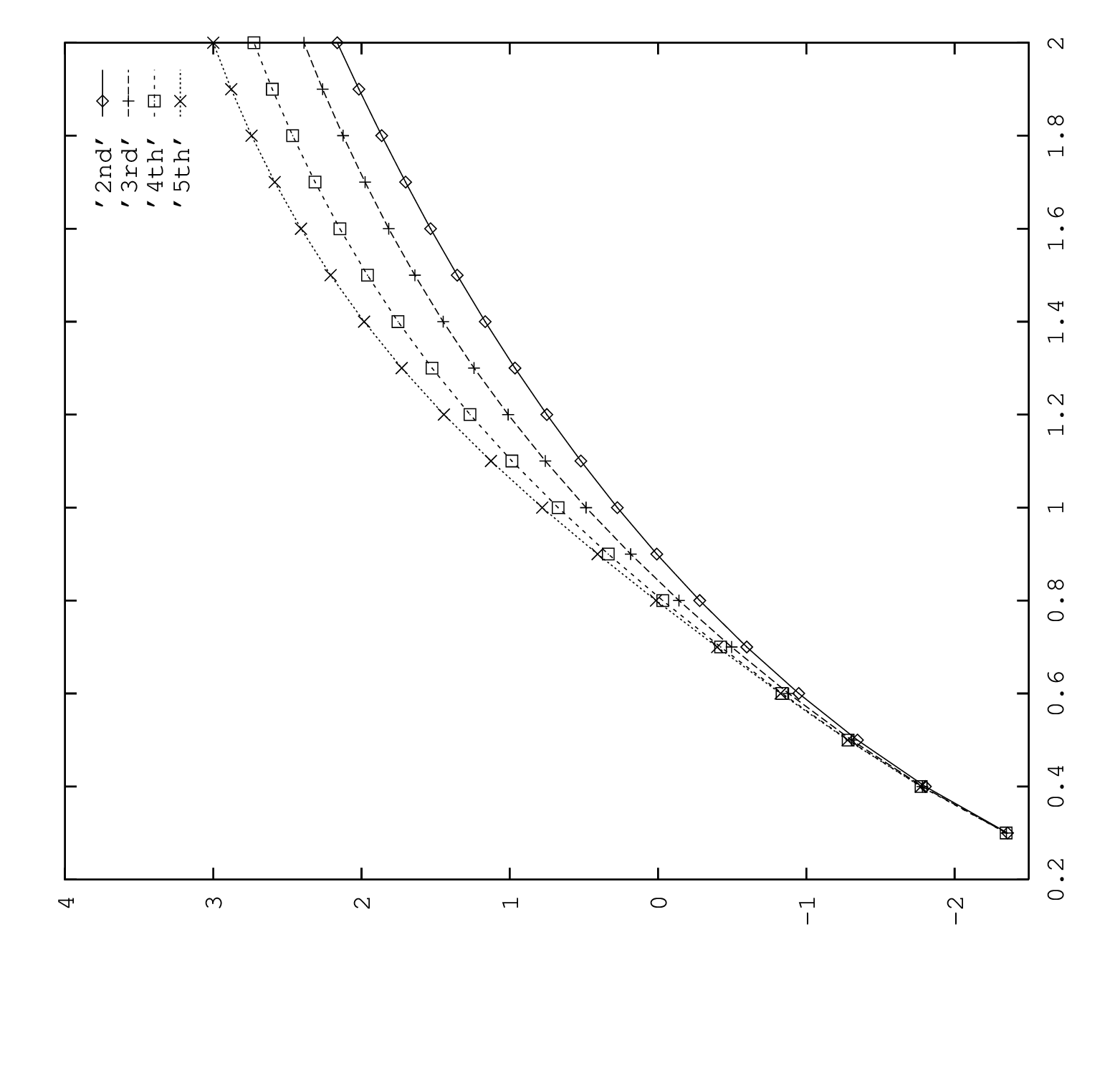}
\vspace{2mm}
\caption{$\mu_2\prime$ as a function of $\beta$}
\label{fig2}
\end{figure}

\begin{figure}[htb]
\fpsxsize=7.5cm
\hspace{-15mm}
\def\fpsangle{270}
\fpsbox[70 90 579 760]{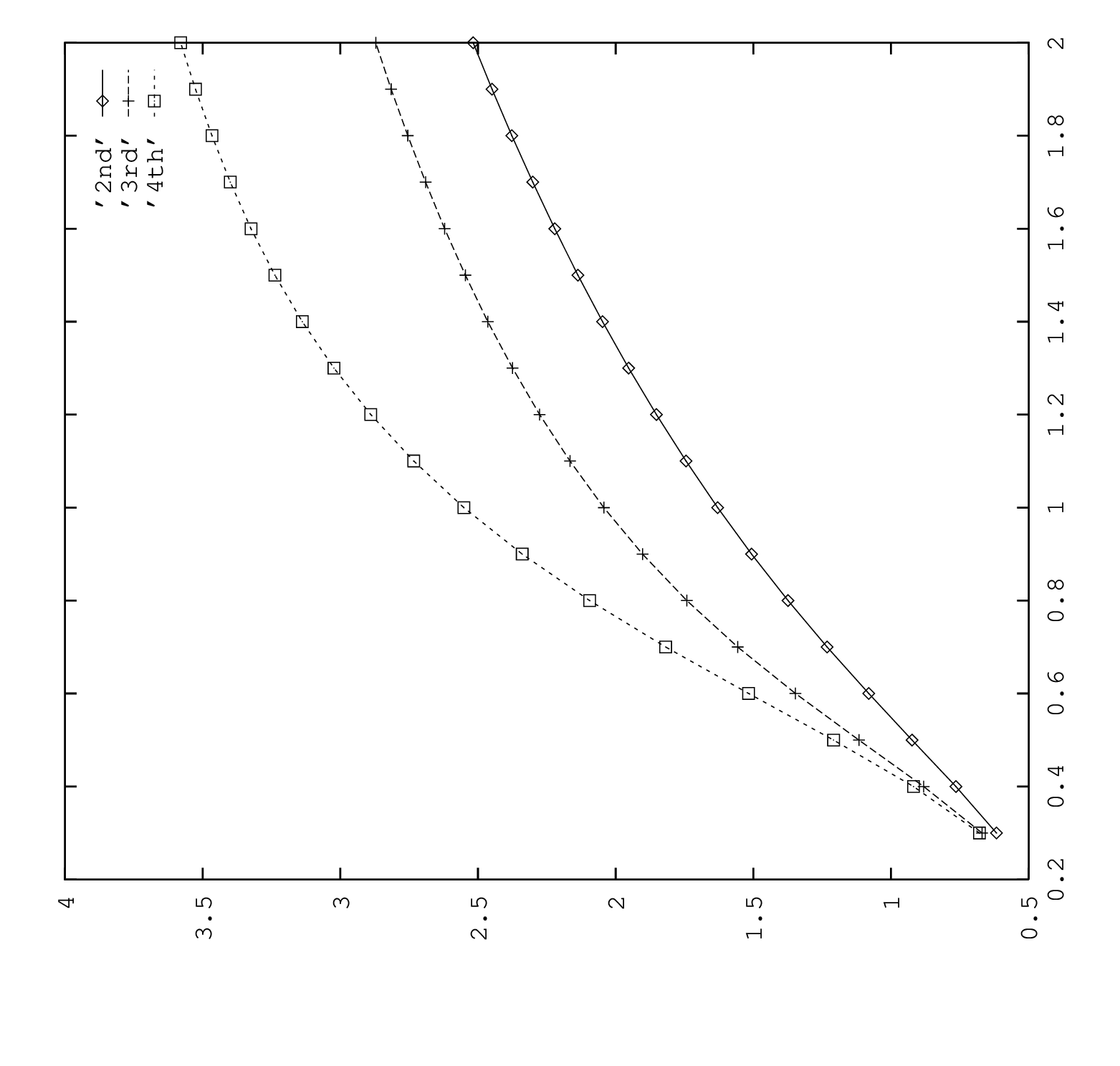}
\vspace{-2mm}
\caption{$\mu_0^{A}\prime$ as a function of $\beta$}
\label{fig3}
\end{figure}

\end{document}